\documentclass{pas}

\usepackage{natbib}
\usepackage{lipsum}

\usepackage{amsmath}
\usepackage{graphicx}%
\usepackage{dcolumn}%
\usepackage{bm}%
\usepackage{cancel} %
\usepackage[dvipsnames]{xcolor} %
\usepackage{multirow}
\usepackage{array}
\usepackage{xurl}

\usepackage{placeins}

\begin{document}

\lefttitle{Publications of the Astronomical Society of Australia}
\righttitle{S. Tiruvaskar and C. Gordon}

\jnlPage{1}{12}
\jnlDoiYr{2026}
\doival{10.1017/pasa.xxxx.xx}

\articletitt{Research Paper}

\title{Replacing Gaussian Processes with Neural Networks in Pulsar Timing Array Inference of the Gravitational-Wave Background}

\author{ \gn{Shreyas} \sn{Tiruvaskar}$^{1}$ and \gn{Chris} \sn{Gordon} $^{1}$}

\affil{$^1$ School of Physical and Chemical Sciences, University of Canterbury, Christchurch, New Zealand}

\corresp{C. Gordon, Email: chris.gordon@canterbury.ac.nz}


\history{(Received xx xx xxxx; revised xx xx xxxx; accepted xx xx xxxx)}

\begin{abstract}
Bayesian inference of nanohertz gravitational-wave background models in pulsar timing array analyses often relies on Gaussian-process interpolators to avoid repeated, computationally expensive strain-spectrum calculations. However, Gaussian-process training becomes a bottleneck for large training sets. We test whether probabilistic neural networks can replace Gaussian processes in this role for a self-interacting dark matter model and a phenomenological environmental model. We find that neural networks recover consistent posteriors while significantly reducing both training and Markov chain Monte Carlo runtime, with the largest gains for the more computationally demanding model.
\end{abstract}


\maketitle

\section{\label{sec:intro}Introduction}

Pulsar timing array (PTA) observations \citep{gwb_nanograv, gwb_ppta, gwb_epta_2023, gwb_cpta_2023} have made the nanohertz gravitational-wave background (GWB) a powerful probe of the cosmic population of supermassive black hole binaries (SMBHBs) and of the astrophysical environments that govern their evolution (e.g.~\cite{gwb_smbh_Begelman_1980, Agazie_2023}). SMBHBs are widely regarded as the leading astrophysical source of the PTA signal, and recent evidence for a common-spectrum stochastic process with inter-pulsar correlations consistent with a gravitational wave (GW) origin has made it increasingly important to connect the measured GWB spectrum to the underlying demographics and dynamics of the binary population. Interpreting PTA data in terms of physical source models is, however, computationally demanding. A Bayesian analysis must evaluate the predicted strain spectrum many times across a multidimensional parameter space, and each direct forward-model calculation can be expensive when the binary dynamics and population modelling are treated in detail.

A practical way to make this inference tractable is to precompute a library of simulated spectra and use an interpolator, or emulator, within the Markov chain Monte Carlo (MCMC) analysis. This strategy has already been adopted in PTA studies through Gaussian-process (GP) interpolation within the \texttt{holodeck} framework \citep{Taylor_2017,Agazie_2023}. In that setting, the GP replaces repeated direct simulations during likelihood evaluation and thereby makes large-scale Bayesian sampling feasible. However, as the size of the training library grows, GP training itself becomes costly. This becomes particularly restrictive when denser sampling of parameter space is required to model more nonlinear signals accurately, or when the underlying astrophysical model contains many free parameters that must be explored simultaneously.

Neural-network and neural-density-estimation methods have also been explored in related PTA and stochastic-background applications, including rapid posterior estimation and SMBHB population emulation \citep{Shih2024,Bonetti2024,Laal2025,Vallisneri2025}. These studies address a range of inference and emulation tasks, including approaches that model more detailed strain distributions. The problem considered here is narrower and deliberately pipeline-focused: we test whether the GP interpolators used in the existing \texttt{holodeck}-based Bayesian inference workflow can be replaced by computationally lightweight probabilistic feedforward NNs without otherwise modifying the likelihood structure.

 The purpose of this work is not to introduce a general-purpose density estimator for SMBHB strain realisations. Instead, we ask a practical question motivated by current PTA inference workflows: can the computationally expensive GP step be replaced by a lightweight probabilistic neural-network (NN) emulator while preserving the posterior constraints obtained from the established summary-statistic likelihood?

We apply this comparison to two astrophysical models. The first is a six-parameter self-interacting dark matter (SIDM) model in which the dark matter (DM) environment affects SMBHB inspiral and hence the resulting GWB spectrum (\cite{Tiruvaskar_2026}; hereafter \hyperlink{cite.Tiruvaskar_2026}{TG2026}). The second is the phenomenological environmental model used in \texttt{holodeck}-based PTA analyses by \cite{Agazie_2023} (hereafter \hyperlink{cite.Agazie_2023}{Agazie2023}). These two cases provide a useful contrast: the SIDM model is more computationally demanding and requires a larger training set, whereas the phenomenological model offers a simpler benchmark with an established GP-based analysis.

Our main result is that NNs can replace GPs in this setting without degrading the inferred astrophysical constraints, while substantially reducing the computational cost. For the larger SIDM training set, the NN interpolators reduce training time by up to nearly two orders of magnitude and also accelerate the subsequent MCMC analysis. For the phenomenological model, the gain is smaller but still significant. The overall outcome is a faster Bayesian pipeline that remains sufficiently accurate for practical PTA parameter inference.

The structure of this paper is as follows. In Sec.~2 we summarise the SIDM model considered in our analysis. In Sec.~3 we describe the interpolation and Bayesian inference methodology, including the GP and NN implementations. In Sec.~4 we apply the same accelerated pipeline to the phenomenological model. In Sec.~5 we compare the performance of the two interpolators in terms of training time, prediction accuracy, and posterior recovery. We conclude in Sec.~6.

\section{Self-interacting dark matter model}
The first model we consider is an astrophysical SIDM halo model for a galaxy hosting a central SMBH.
 In this model, the SIDM density profile exhibits a spike near the centre (\cite{alonso}; hereafter \hyperlink{cite.alonso}{ACD2024}).

We consider a merger of two galaxies with SMBHs at their centres embedded in spherically symmetric SIDM halos. After these DM halos and the stellar contents of the galaxies merge, the two central black holes form a binary, which inspirals and eventually merges. During this process, the SMBHB emits GWs. As the SMBHB moves through the SIDM halo, the dynamical friction provided by DM removes kinetic energy from the SMBHB. If the SIDM halo can extract sufficient kinetic energy from the SMBHB without getting disrupted, it can make the black holes merge within the age of the universe, solving the ``final parsec problem'' (\hyperlink{cite.alonso}{ACD2024}).

Since this model predicts GW emission during the SMBHB merger, we can use the observed PTA GWB data to constrain its parameters. \hyperlink{cite.Tiruvaskar_2026}{TG2026} did this by performing Bayesian inference with MCMC using NANOGrav PTA GWB data.
 The resulting posterior distributions of the model parameters were presented by \hyperlink{cite.Tiruvaskar_2026}{TG2026}. To perform this statistical analysis, we used a Python package developed by the NANOGrav collaboration, known as \texttt{holodeck}, which is described by \hyperlink{cite.Agazie_2023}{Agazie2023}.

For this analysis, we begin after the SIDM halos of the merging galaxies have coalesced into a single halo. We then simulate the inspiral and merger of the SMBHB within this halo and compute the associated gravitational-wave emission. The resulting characteristic-strain spectrum is calculated for each merger, and the spectra from the full SMBHB population are combined to obtain the total GWB.

To simulate the SMBHB inspiral and merger, we need to know the dynamics of the system, which can be determined by calculating the gravitational force and the dynamical friction. The dynamical friction experienced by the SMBH can be calculated if we know the DM density around it. As the SMBHs inspiral toward the centre of the halo, they move through regions of increasing SIDM density. Thus, to calculate the dynamics of the binary merger, we first need to determine the SIDM density profile.

In the innermost region of the DM halo, the density is
predicted to follow a power-law spike profile 
(\hyperlink{cite.alonso}{ACD2024}),
\begin{equation}
    \rho_{\mathrm{sp}}(r) = \rho_{\mathrm{sp}0} \left( \frac{r_{\mathrm{sp}}}{r} \right)^\gamma,
\label{spike}
\end{equation}
where \(\rho_{\mathrm{sp}0}\) is the constant scaling density, \(r_{\mathrm{sp}}\) is the spike radius and $r$ is the radius from the centre of the galaxy.

 In the spike region, the exponent, \(\gamma\), from Eq. \eqref{spike} can have different values depending on the type of interactions between the DM particles. For example, if we have contact interactions, \(\gamma=3/4\), and for the Coulomb interactions, \(\gamma=7/4\) (\hyperlink{cite.alonso}{ACD2024}). Inside the spike region, the velocity dispersion of SIDM particles \(v(r)\) increases towards the centre of the galaxy. Following \hyperlink{cite.alonso}{ACD2024}, we considered interactions that have a massive mediator as the force carrier. For this type of interaction, the exponent $\gamma$ is not constant throughout the spike region, but it changes from 3/4 to 7/4 as  \(v(r)\) becomes greater than the transition velocity \(v_t\). We make this transition velocity a free parameter of our model.

The other free parameter relating to SIDM particles is \(
\left({\sigma_0}/{m}\right)\left({t_{\mathrm{age}}}/{1\,\mathrm{Gyr}}\right),
\) where \(\sigma_0\) 
denotes the low-velocity normalisation of the SIDM self-interaction cross section $\sigma$, \(m\) is the SIDM particle mass, and \(t_\mathrm{age}\) is the age of the DM isothermal core (\hyperlink{cite.alonso}{ACD2024}). 
Although the MCMC sampling is performed in terms of \(
\left({\sigma_0}/{m}\right)\left({t_{\mathrm{age}}}/{1\,\mathrm{Gyr}}\right)
\),
for comparison with previous work, we present the
corresponding posterior samples for \(\sigma/m\). This
transformation is applied only in post-processing and does not
affect the sampled parameterization of the MCMC. Following
\hyperlink{cite.alonso}{ACD2024} and
\hyperlink{cite.Tiruvaskar_2026}{TG2026}, this transformation requires choosing typical values for
the total binary mass, the binary mass ratio, and the merger
redshift. We take these benchmark values to be
\((1.59\times10^{9}\,M_\odot,\ 0.88,\ 0.87)\), respectively.
These values are chosen because SMBHBs in this region of
parameter space make a significant contribution to the GWB
signal of the SIDM model in the lowest PTA frequency band
(\hyperlink{cite.Tiruvaskar_2026}{TG2026}).

Once the SIDM density profile is computed, we compute the GW signal for a binary merger in that halo. To calculate the GWB and compare it with the PTA data, we add all the strain spectra caused by all the SMBHBs in a specific mass and redshift range. This is done by calculating the differential number density of galaxy mergers, which is expressed in terms of the galaxy stellar mass function (GSMF). Two parameters from GSMF, \(\psi_0\) and \(m_{\psi,0}\), are also varied in our analysis following \hyperlink{cite.Agazie_2023}{Agazie2023}.


The last two parameters we vary are \(\mu\) and \(\epsilon_\mu\),
which describe the relation between SMBH mass and the stellar
bulge mass of the host galaxy. In the notation of
\hyperlink{cite.Agazie_2023}{Agazie2023}, \(\mu\) is the
normalisation of this relation, and \(\epsilon_\mu\) is its intrinsic
scatter. For compactness, we label this relation as MMB in the tables.


These are the six model parameters that we vary in our analysis. Details about each of them are given by \hyperlink{cite.Tiruvaskar_2026}{TG2026}. We also summarise these parameters and their priors in Table \ref{table:sidm_params}.

\begin{table}[htbp]
    \centering
    {
    \renewcommand{\arraystretch}{1.5}
    \begin{tabular}{|c|c|}
        \hline
        \textbf{Model parameter} & \textbf{Prior} \\
        \hline
        $v_t$        & $\mathcal{U}(1, 2000)$ km/s \\
        $\frac{\sigma_0}{m}\frac{t_\mathrm{age}}{1\,\mathrm{Gyr}}$   & $\mathcal{U}(0.01, 200)$ $\mathrm{cm^2/g}$ \\
        \hline
       GSMF $\psi_0$        & $\mathcal{N}(-2.56, 0.4)$ \\
        GSMF $m_{\psi,0}$    & $\mathcal{N}(10.9, 0.4)$ \\
        \hline
        MMB $\mu$           & $\mathcal{N}(8.6, 0.2)$ \\
        MMB $\epsilon_\mu$  & $\mathcal{N}(0.32, 0.15)$ dex \\
        \hline
    \end{tabular}
    }
\caption{Priors for the SIDM model parameters. We denote uniform distributions with $\mathcal{U}$(min,max) and
Gaussian distributions with $\mathcal{N}$(mean, std.~dev.).}
\label{table:sidm_params}
\end{table}

\section{\label{sec:methods}Methods}

Given the simulated strain spectra from \texttt{holodeck}, we generate a library of spectra over the model parameter space.
This library forms the training set for the interpolator. Then, we train a GP on this training dataset so that it can interpolate at any parameter values and predict strain spectra for that parameter combination. 

For each PTA frequency bin \(f_i\), we characterize the GWB  by 
\begin{equation}
x_i\equiv \log_{10}\!\left(h_c^2(f_i)\right)\, .
\end{equation}
where $h_c$ is the characteristic strain (\hyperlink{cite.Agazie_2023}{Agazie2023}).
When we generate the library, for each parameter combination, we generate \(R\) realisations of the strain-spectrum for each of the 5 lowest PTA frequencies. Following \hyperlink{cite.Agazie_2023}{Agazie2023}, we set the number of strain-spectrum realisations, for a particular parameter combination, to be \(R=2000\). Next, for each of the five frequency bins, we train two GPs, one on the median values of $x_i$ and the other on the standard deviations of $x_i$. Thus, the trained GPs can predict the median and the standard deviation of $x_i$ for any values in the parameter space. For each frequency bin, the GP trained on the median values will give a predicted value of the median and an uncertainty in its prediction of the median. Similarly, for the standard deviation.

Following \hyperlink{cite.Agazie_2023}{Agazie2023}, we model the probability distribution $p\left(x_i \mid \Theta\right)$ as a Gaussian with $\Theta$ denoting the astrophysical parameters. The adequacy of this approximation is supported by the high validation accuracy of the GP reconstruction and by the near-equivalence of the resulting posteriors on $\Theta$ to those from the full timing-residual likelihood (\hyperlink{cite.Agazie_2023}{Agazie2023}). Thus, the effective Gaussian distribution for the model prediction in frequency bin \(i\) is centred on the interpolator-predicted value and has variance
\begin{equation}
\sigma_{\mathrm{total},i}^2
=
\left(\sigma_{\mathrm{median},i}^{\mathrm{pred}}\right)^2
+
\left(\sigma_{\mathrm{std},i}^{\mathrm{pred}}\right)^2
+
\left(\mathrm{std}_i^{(\mathrm{pred})}\right)^2\, .
\label{eq:gp_total_sigma}
\end{equation}
This variance calculation comprises the variance in the median prediction $\left(\sigma_{\mathrm{median},i}^{\mathrm{pred}}\right)^2$, the variance in the standard deviation prediction $\left(\sigma_{\mathrm{std},i}^{\mathrm{pred}}\right)^2$, and the predicted standard deviation itself $\left(\mathrm{std}_i^{(\mathrm{pred})}\right)^2$.

 A more general emulator could instead attempt to learn the full strain-realization distribution. That is a different objective from the one pursued here. Our aim is to determine whether the existing median-and-standard-deviation representation, already validated for the \texttt{holodeck}-based PTA analysis, can be accelerated without changing the likelihood model or shifting the inferred astrophysical posteriors.

\hyperlink{cite.Agazie_2023}{Agazie2023} (in their Sec.~3.5) explain how a transformed version of $p\left(x_i \mid \Theta\right)$ is used in calculating the likelihood of the PTA data, which is then combined with the priors on $\Theta$ to obtain the posterior distribution of  \(\Theta\) given the PTA data. The MCMC procedure samples this posterior distribution, thereby providing the inferred constraints on \(\Theta\).

\subsection{\label{subsec:gp_training}Gaussian processes}
GPs have long been used in spatial statistics and geostatistics (e.g.~\cite{noel_gardar_2008}). A GP can be used to interpolate between the known data points and predict a distribution at intermediate points (e.g.~\cite{Rasmussen1997EvaluationOG, williams1996gp,ambikasaran2015}).

Agazie2023 outlined the use of GPs in PTA GWB analyses.
We use the \texttt{George} GP regression library
\cite{ambikasaran2015} for GP training. While the
computational cost of GP training scales as
\(\mathcal{O}(N^3)\) with the number of training points
\(N\) \citep{noack2022}, \texttt{George} implements more
efficient algorithms that speed up this procedure. Further
details can be found in the \texttt{George} documentation.

We generated a training library of 8000 parameter points and
the corresponding strain spectra, as described by
\hyperlink{cite.Tiruvaskar_2026}{TG2026}. We then trained two
separate GPs, one on the medians of \(x_i\) and the other on
their standard deviations. This training was carried out
independently for the five lowest PTA frequency bins.

Because GP training becomes increasingly expensive as the size
of the training set grows, we first tested whether a smaller
value of \(N\) would be sufficient. Increasing the training set
from 2000 to 8000 points raises the GP training time from
\(\sim 2.3\) hours to \(\sim 33\) hours. All training and timing
measurements reported in this article were obtained on the University of
Canterbury Research Cluster using a 200-core CPU node with an
AMD EPYC-Milan processor. Since this increase represents a
substantial overhead in the statistical-analysis pipeline, we
followed \hyperlink{cite.Agazie_2023}{Agazie2023} and began
with a training library of 2000 points.

As a targeted check of the GP interpolator near the region most
relevant for the inference, we generated MCMC samples using the
GPs trained on this 2000-point library. At the maximum-posterior
parameter point, we computed \(x_i\) directly using  \texttt{holodeck}
and compared the result with the corresponding GP prediction.

We define the interpolator as passing this maximum-posterior
validation check if, at the same parameter values, the median from
the direct  \texttt{holodeck} simulations lies within the interpolator's
95\% predictive interval in each of the five frequency bins. This is a
local validation check at the posterior mode, rather than a global
validation of the interpolator over the full prior volume.

\begin{figure}[h!]
  \centering
  \includegraphics[width=0.45\textwidth]{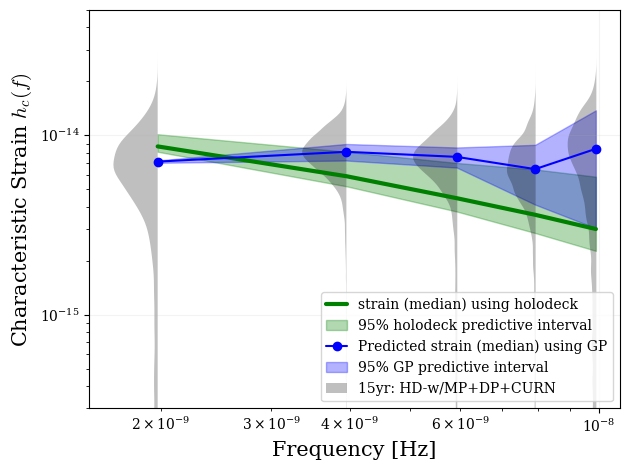}
  \includegraphics[width=0.45\textwidth]{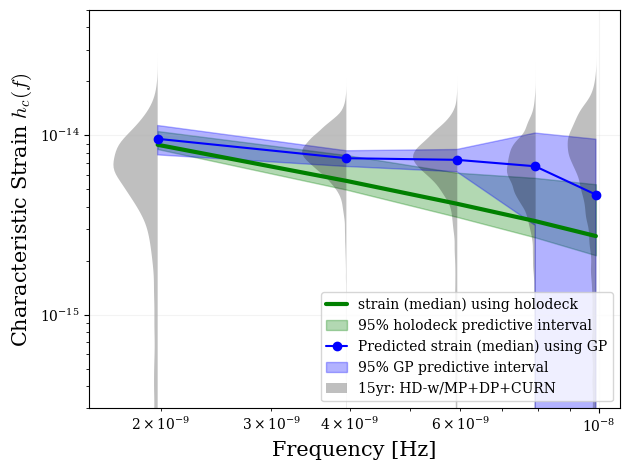}
   \includegraphics[width=0.45\textwidth]{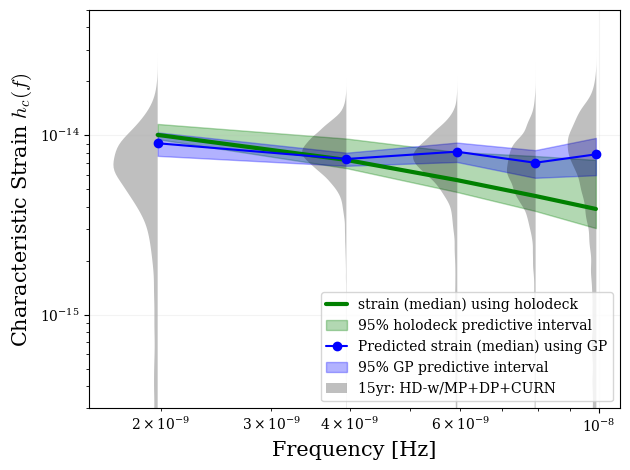}

  \caption{Comparison of GP predictions with direct \texttt{holodeck}
simulations for the SIDM model at the maximum-posterior parameter values
obtained from MCMC runs using GPs trained on libraries of 2000, 4000,
and 8000 points. The top, middle, and bottom panels correspond to the
2000-, 4000-, and 8000-point training libraries, respectively. In each
panel, the solid blue curve shows the GP-predicted median
characteristic-strain spectrum, and the blue shaded region shows the
nominal 95\% predictive interval. The solid green curve shows the median
spectrum from 2000 direct \texttt{holodeck} realisations at the same
parameter values, and the green shaded region shows the corresponding
central 95\% range. The agreement improves as the GP training library is
enlarged, but the GP predictive intervals do not consistently contain
the \texttt{holodeck} median across all five frequency bins.}
\label{fig:gp_holodeck_strain_sidm}

\end{figure}

In Fig.~\ref{fig:gp_holodeck_strain_sidm}, we fix the model
parameters to their maximum-posterior values, obtained from the MCMC chain, and plot the
resulting predictive distributions for the characteristic-strain
spectrum. The blue lines and shaded regions show the GP
prediction for this parameter combination. The solid blue lines
represent the GP-predicted median, denoted as
\(\mathrm{median}^{(\mathrm{pred})}\). The blue shaded region shows
the nominal 95\% predictive interval for the GP, computed using the total
uncertainty from Eq.~\ref{eq:gp_total_sigma}. The green lines
and shaded regions show the strain spectra produced by
\texttt{holodeck} simulations for the same maximum-posterior
parameter combination. As described above, we generate 2000
\texttt{holodeck} realisations for each spectrum. The solid green
lines represent the median values of these realisations, and the
shaded green region spans the 2.5th to 97.5th percentiles,
corresponding to the central 95\% range of characteristic
strains.

Under this maximum-posterior validation criterion, the SIDM GP
interpolators do not pass the check shown in Fig.~\ref{fig:gp_holodeck_strain_sidm}. Although increasing the training library from 2000 to 8000 points improves the agreement, the  \texttt{holodeck} median is still not consistently contained within the GP 95\% predictive interval across all five frequency bins.

\begin{figure*}[h!]
  \centering
  \includegraphics[width=0.9\textwidth]{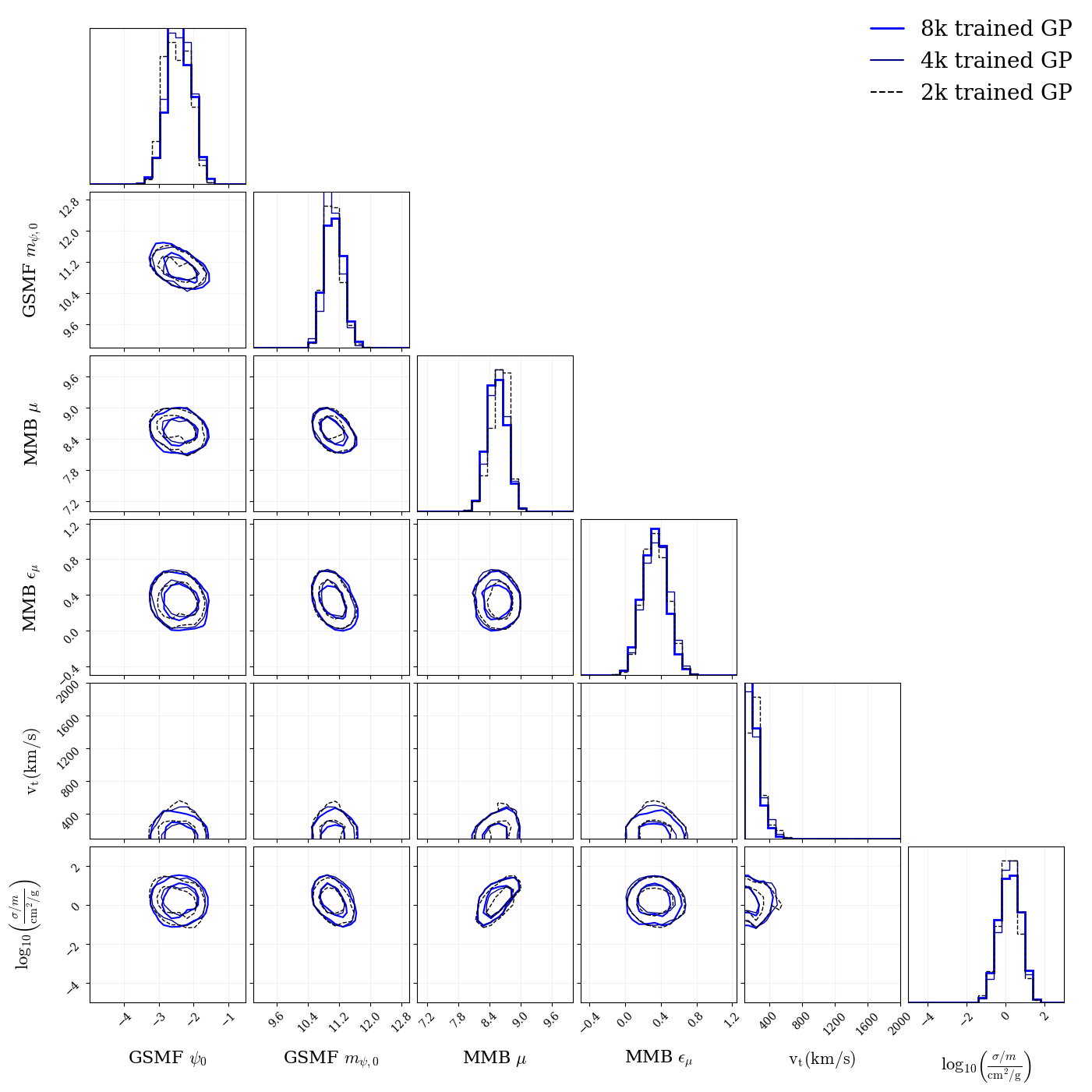}
   \caption{Corner plot of the posterior distributions for the SIDM model parameters from MCMC runs using GP interpolators trained on 2000, 4000, and 8000 library points. The contours denote the 68\% and 95\% credible regions.}
  \label{fig:mcmc_gp_2k_4k_8k}
\end{figure*}

The posterior distribution of the model parameters obtained from an
MCMC analysis using GPs trained on 8000 training points was also
calculated by \hyperlink{cite.Tiruvaskar_2026}{TG2026}. In this article,
we show this result as the blue contours in
Fig.~\ref{fig:mcmc_gp_2k_4k_8k}. For comparison, we also show the
posteriors from MCMC analyses using GPs trained on 2000 and 4000
training points in Fig.~\ref{fig:mcmc_gp_2k_4k_8k}. We observe no
significant differences in the posterior distributions across the three
training-set sizes. However, the maximum-posterior spectrum comparison
in Fig.~\ref{fig:gp_holodeck_strain_sidm} shows that increasing the
training set improves the agreement with the direct  \texttt{holodeck}
simulations, although the 8000-point GP still does not fully satisfy our
median-containment validation criterion. We therefore use the 8000-point
case as the strongest available GP baseline for the subsequent
comparison. The posterior medians and the 16th and 84th percentiles for
the three GP analyses are reported in
Table~\ref{tab:sidm_posterior_median_gp}.

\begin{table}
\centering
\renewcommand{\arraystretch}{1.6}
\begin{tabular}{|l|r|r|r|}
\hline
Parameter & 8000 & 4000 & 2000 \\
\hline
GSMF $\psi_0$ & $-2.43^{+0.38}_{-0.31}$ & $-2.42^{+0.36}_{-0.35}$ & $-2.50^{+0.40}_{-0.36}$ \\
GSMF $m_{\psi, 0}$ & $11.05^{+0.25}_{-0.23}$ & $11.01^{+0.24}_{-0.16}$ & $11.02^{+0.24}_{-0.20}$ \\
MMB $\mu$ & $8.53^{+0.19}_{-0.18}$ & $8.55^{+0.18}_{-0.18}$ & $8.61^{+0.14}_{-0.21}$ \\
MMB $\epsilon_{\mu}$ & $0.33^{+0.13}_{-0.14}$ & $0.35^{+0.14}_{-0.15}$ & $0.33^{+0.15}_{-0.13}$ \\
$\mathrm{v_t (km/s)}$ & $183.37^{+93.50}_{-74.86}$ & $189.97^{+118.02}_{-78.85}$ & $222.55^{+86.85}_{-83.38}$ \\
$\mathrm{log}_{10} \left( \frac{\sigma / m}{\mathrm{cm}^2 /\mathrm{g}} \right)$ & $0.22^{+0.56}_{-0.57}$ & $0.25^{+0.50}_{-0.52}$ & $0.18^{+0.50}_{-0.53}$ \\
\hline
\end{tabular}
\caption{Parameter constraints from MCMC analyses using GPs trained on 8000, 4000, and 2000 training points. For each parameter, we list the posterior median and the 16th and 84th percentiles, which define the central 68\% credible interval.}

\label{tab:sidm_posterior_median_gp}
\end{table}

\subsection{Neural networks}

\subsubsection{Principle and architecture}

A deep NN consists of multiple layers of
interconnected neurons, each of which applies a linear
transformation followed by a nonlinear activation function to its
input. Given an input vector, the data are propagated forward
through successive layers. In each layer, the input is transformed
using trainable weight matrices and bias vectors, and
the resulting activations are passed to the next layer. This
process continues until the output layer produces the final
prediction. See \cite{goodfellow2016deep}, for example, for further details of this formalism.

At the output layer, the network prediction is compared with the
corresponding target values from the training set. Their
difference defines the loss, \(\mathcal{L}\).

\subsubsection{Hyperparameter selection}\label{subsub:hyperparam_nn}
In our implementation, for the network trained on the median values, we employed three hidden layers containing 16, 32, and 16 neurons, respectively. We initially tested architectures with fewer layers and nodes, progressively increasing the network complexity until the training and validation losses no longer showed significant improvement without the onset of overfitting. Overfitting occurs when the NN becomes overly specialised to the training data and fails to generalise effectively to the validation set, resulting in a validation loss significantly larger than the training loss. Consequently, architectures with too few layers and nodes produced comparatively larger training and validation losses, whereas architectures with too many layers and nodes led to overfitting.

We adopted the ReLU (rectified linear unit) activation function, \(f(x)=\max(0,x)\), which mitigates the vanishing gradient problem, thereby enabling deep neural networks to converge more rapidly and learn more effectively, as discussed by \cite{nair2010relu}. We minimised the loss using the Adam optimiser, as it combines the advantages of Momentum (faster convergence) and RMSprop (adaptive learning rates), as described by \cite{kingma2015adam}.

\subsubsection{Training}
As mentioned before, the training dataset is simulated using \texttt{holodeck}. In our training set, we chose 8000 parameter combinations using the Latin hypercube sampling as explained by \cite{taylor_lhc} and by \hyperlink{cite.Agazie_2023}{Agazie2023}. This procedure yields parameter combinations that uniformly span our six-dimensional parameter space. For each such combination, we simulate the  $x_i$ in the five lowest PTA frequency bins, producing 2000 realisations of the $x_i$ for that parameter point. This is explained in detail by \hyperlink{cite.Tiruvaskar_2026}{TG2026} in their Sec.~V(C). 
Following the convention adopted in Sec. 3.4 of \hyperlink{cite.Agazie_2023}{Agazie2023},
we assign the GP/NN training target for the median of \(x_i\) an input
uncertainty
\begin{equation}
    \sigma_{\mathrm{median},i}^{\mathrm{holo}}
    =
    \frac{\mathrm{std}^{\mathrm{holo}}_i}{\sqrt{R}},
    \label{eq:median-error}
\end{equation}
where \(\mathrm{std}^{\mathrm{holo}}_i\) is the standard deviation of
\(x_i\) over the \(R\)  \texttt{holodeck} realisations at frequency bin
\(i\). Similarly, we assign the training target for the standard
deviation of \(x_i\) the input uncertainty
\begin{equation}
    \sigma_{\mathrm{std},i}^{\mathrm{holo}}
    =
    \frac{\mathrm{std}^{\mathrm{holo}}_i}{\sqrt{2(R - 1)}} .
    \label{eq:std-error}
\end{equation}
At each frequency bin, two GPs are trained: one on the median of $x_i$ and one on its standard deviation. For any new point in parameter space, these GPs return the predicted median and standard deviation, together with their respective interpolation uncertainties. We do the same while training our NNs.

Our training input consists of 8000 parameter combinations across our 6 model parameters. The shape of the training input is (8000, 6). 
Our \texttt{holodeck} simulation values are shaped (8000, 10) as we combine \texttt{holodeck} simulation $x_i$ median values (which we denote as $\mathrm{median}^{(\mathrm{holo})}$) for each of the 5 frequency bins, and the corresponding $\sigma_{\mathrm{median}}^{(\mathrm{holo})}$, which are given by Eq.~\ref{eq:median-error}.

We use a probabilistic NN, rather than a deterministic NN, so that the interpolator returns both the predicted median strain and its predictive uncertainty at each frequency bin.
Therefore, our loss function should not only consider the \texttt{holodeck}-simulated and predicted median values, but also their uncertainties.
Following Secs.~6.2.1.1 and 6.2.2.1 of the book by \cite{goodfellow2016deep}, we train the NN by minimising the negative conditional log-likelihood of the training data. For a Gaussian predictive distribution, this reduces to the following Gaussian negative log-likelihood:

\begin{equation}
    \begin{aligned}
        - &\mathrm{log}\, p( \mathrm{median}^{(\mathrm{holo})}|\mathrm{median}^{(\mathrm{pred})}, \sigma_\mathrm{comb}) \\
        & = \frac{1}{2} \left( \frac{(\mathrm{median}^{(\mathrm{holo})}-\mathrm{median}^{(\mathrm{pred})})^2}{\sigma_\mathrm{comb}^2} + \log \left(2\pi \sigma_\mathrm{comb}^2 \right) \right)
    \end{aligned}
\end{equation}
where the combined uncertainty $\sigma_\mathrm{comb}$ is the quadrature sum of the uncertainty in prediction, \(\sigma_\mathrm{median}^{(\mathrm{pred})}\), and the uncertainty in \texttt{holodeck} simulation median value, \(\sigma_{\mathrm{median}}^{(\mathrm{holo})}\):
\begin{equation}
    \sigma_\mathrm{comb} = \sqrt{\left(\sigma_\mathrm{median}^{\mathrm{pred}} \right) ^2 + \left(\sigma_{\mathrm{median}}^{\mathrm{holo}} \right)^2}\,.
\end{equation}
Here $\mathrm{median}^{(\mathrm{pred})}$ and $\sigma_\mathrm{median}^{(\mathrm{pred})}$ are outputs of the NN.
This negative log-likelihood is calculated for five median values corresponding to the five frequency bins. Summing these five negative log-likelihood values, we get our loss function as
\begin{equation}
\mathcal{L}=\sum_{i=1}^5-\log p\left(\operatorname{median}_i^{(\text{holo })} \mid \operatorname{median}_i^{(\text{pred) }}, \sigma_{\mathrm{comb},i}\right)\,.
\end{equation}

We train the NN to minimise the $\mathcal{L}$. We generate another dataset of 8000 points for test and validation. We use 4000 data points from this independently generated dataset as a validation set and the other 4000 as the test set. At each epoch during the training, in addition to the loss for the training data, the loss is calculated for the validation data as well. We use the test set when generating Fig.~\ref{fig:gp_nn_error_sidm}.

In our case, training was performed for 1000 epochs with early stopping based on validation loss and patience=100, which means that if the validation loss does not decrease for 100 epochs, training stops. The NN parameters corresponding to the minimum validation loss were automatically restored and used for subsequent analysis. 
Our probabilistic NN was implemented using \texttt{Keras}
\citep{keras}, an application programming interface (API) built on
top of \texttt{TensorFlow}. \texttt{TensorFlow}
\citep{tensorflow} is a Python-based platform for machine
learning. In this work, we used \texttt{TensorFlow} version 2.15.1
and \texttt{tensorflow\_probability} version 0.23.0.

We performed a similar process with the same architecture to train an NN for the standard deviations of the strain spectra. In the training input, instead of median values, we use standard deviations, and instead of median sampling uncertainties, we use sampling uncertainties in standard deviation, which are given in Eq.~\ref{eq:std-error}. Everything else is identical. 

The total training time for both NNs was 13.4 minutes. This is almost 150 times faster than the GP training time, which was 1976.5 minutes. We report all the training times in Table \ref{table:time}.

Once the NNs are trained, they provide the following quantities: \(\mathrm{median}^{(\mathrm{pred})}\), \(\mathrm{std}^{(\mathrm{pred})}\), \(\sigma_\mathrm{median}^\mathrm{pred}\), and \(\sigma_\mathrm{std}^\mathrm{pred}\). These predicted quantities are used, along with the observed PTA data, to calculate the likelihood for MCMC sampling.

\begin{figure}[h!]
  \centering
  \includegraphics[width=0.45\textwidth]{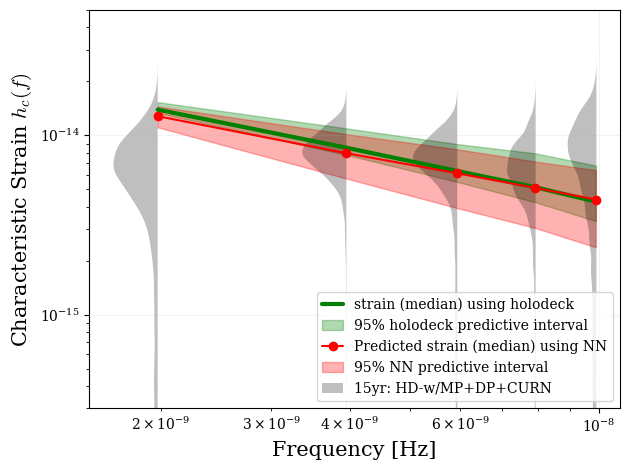}
  \includegraphics[width=0.45\textwidth]{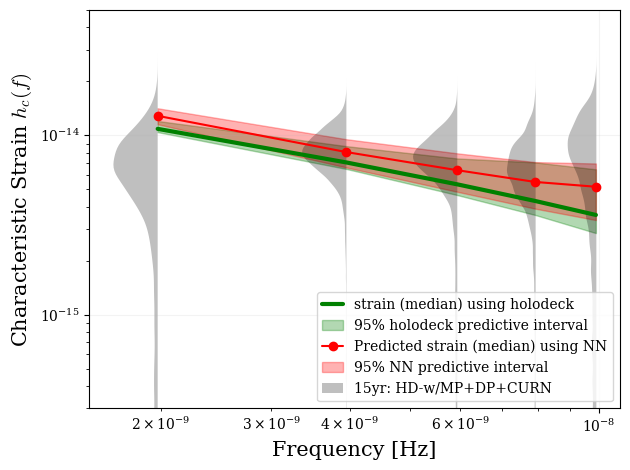}
  \includegraphics[width=0.45\textwidth]{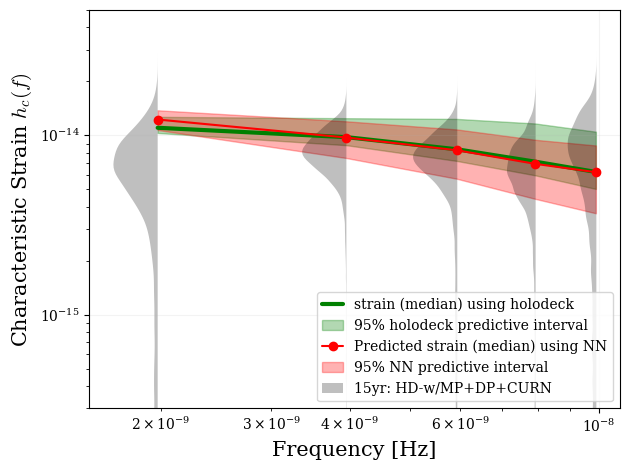}
  \caption{Comparison of NN predictions with direct \texttt{holodeck}
simulations for the SIDM model, using the same plotting convention and
maximum-posterior validation criterion as in
Fig.~\ref{fig:gp_holodeck_strain_sidm}. The maximum-posterior parameter
values are obtained from MCMC runs using NN interpolators. The top,
middle, and bottom panels correspond to NNs trained on 2000, 4000, and
8000 points, respectively. In each panel, the solid red curve shows the
NN-predicted median characteristic-strain spectrum, and the red shaded
region shows the nominal 95\% predictive interval. The solid green curve
shows the median spectrum from 2000 direct \texttt{holodeck}
realisations at the same parameter values, and the green shaded region
shows the corresponding central 95\% range. In contrast to the GP case,
the NN predictive intervals contain the \texttt{holodeck} median across
all five frequency bins for each training-set size considered.}
\label{fig:nn_holodeck_strain_sidm}

\end{figure}

We perform the same maximum-posterior validation check for the
NN interpolators. We fix the model parameters to their
maximum-posterior values from the NN-based MCMC and compare
the resulting NN predictive distributions with the strain spectra
computed directly with  \texttt{holodeck}, as shown in Fig.~\ref{fig:nn_holodeck_strain_sidm}. 
Following the same plotting convention as
Fig.~\ref{fig:gp_holodeck_strain_sidm}, the solid red lines denote the
NN-predicted median characteristic-strain spectra, while the red shaded
regions show the nominal 95\% predictive intervals. These intervals are
computed using the total uncertainty defined analogously to
Eq.~\ref{eq:gp_total_sigma}, with the GP predictive quantities replaced
by the corresponding NN predictive quantities.
The solid green lines show the median values of
the 2000 \texttt{holodeck} realisations, and the green shaded
regions span the 2.5th to 97.5th percentiles of those
realisations, corresponding to the central 95\% range of
characteristic strains.

In contrast to Fig.~\ref{fig:gp_holodeck_strain_sidm}, Fig.~\ref{fig:nn_holodeck_strain_sidm} shows that the NN interpolator passes
this SIDM maximum-posterior validation check. The  \texttt{holodeck}
median lies within the NN 95\% predictive interval in all five
frequency bins. This already holds for the 2000-point NN training library, whereas the corresponding GP validation shown in Fig.~\ref {fig:gp_holodeck_strain_sidm} does not pass this median-containment test.

\begin{figure*}[h!]
  \centering
  \includegraphics[width=0.9\textwidth]{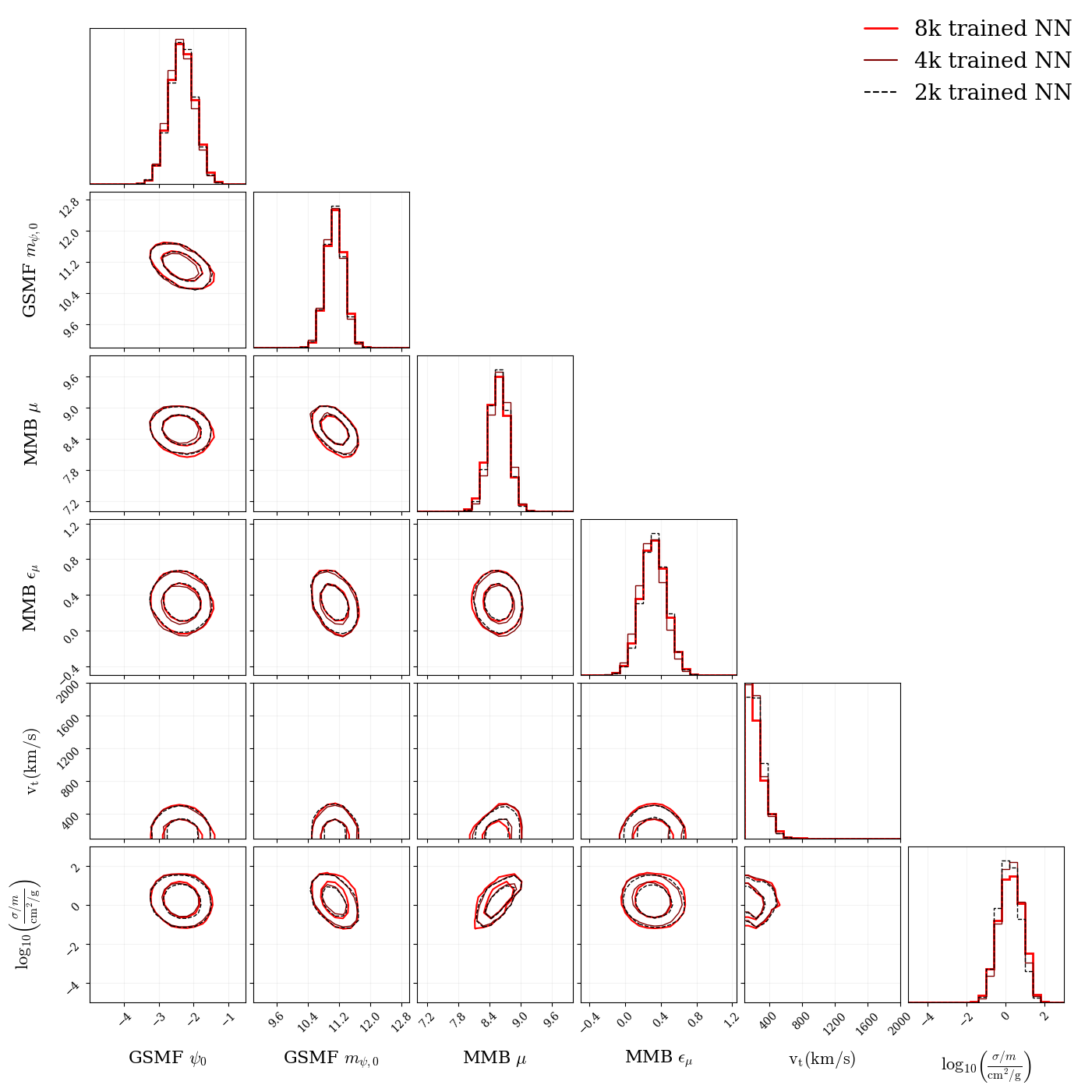}
  \caption{Corner plot of the posterior distributions for the SIDM model parameters from MCMC runs using NN interpolators trained on 2000, 4000, and 8000 library points. The contours denote the 68\% and 95\% credible regions.}
  \label{fig:mcmc_nn_2k_4k_8k}
\end{figure*}

The posterior distributions obtained with NNs are shown in Fig.~\ref{fig:mcmc_nn_2k_4k_8k}. The median values of the NN-based posteriors, together with their 16th and 84th percentiles, are listed in Table~\ref{tab:sidm_posterior_median_nn}.

\begin{table*}
\centering
\renewcommand{\arraystretch}{1.6}
\begin{tabular}{|l|r|r|r|}
\hline
Parameter & 8000 & 4000 & 2000 \\
\hline
GSMF $\psi_0$ & $-2.34^{+0.37}_{-0.36}$ & $-2.37^{+0.36}_{-0.36}$ & $-2.34^{+0.35}_{-0.36}$ \\
GSMF $m_{\psi, 0}$ & $11.09^{+0.24}_{-0.23}$ & $11.07^{+0.23}_{-0.23}$ & $11.09^{+0.23}_{-0.23}$ \\
MMB $\mu$ & $8.56^{+0.19}_{-0.19}$ & $8.59^{+0.18}_{-0.18}$ & $8.57^{+0.18}_{-0.18}$ \\
MMB $\epsilon_{\mu}$ & $0.31^{+0.14}_{-0.14}$ & $0.29^{+0.14}_{-0.14}$ & $0.32^{+0.14}_{-0.14}$ \\
$\mathrm{v_t (km/s)}$ & $184.37^{+130.28}_{-83.03}$ & $210.70^{+107.71}_{-82.02}$ & $217.18^{+113.02}_{-90.96}$ \\
$\mathrm{log}_{10} \left( \frac{\sigma / m}{\mathrm{cm}^2 /\mathrm{g}} \right)$ & $0.26^{+0.62}_{-0.61}$ & $0.26^{+0.56}_{-0.56}$ & $0.17^{+0.57}_{-0.54}$ \\
\hline
\end{tabular}
\caption{Posterior constraints on the SIDM model parameters obtained from MCMC runs using NN interpolators trained on 8000, 4000, and 2000 library points. Listed for each parameter are the median posterior values and the corresponding 16th and 84th percentiles, allowing a direct comparison of the inferred constraints as the NN training-set size is varied.}
\label{tab:sidm_posterior_median_nn}
\end{table*}

\begin{figure*}[htb]
  \centering
  \includegraphics[width=0.9\textwidth]{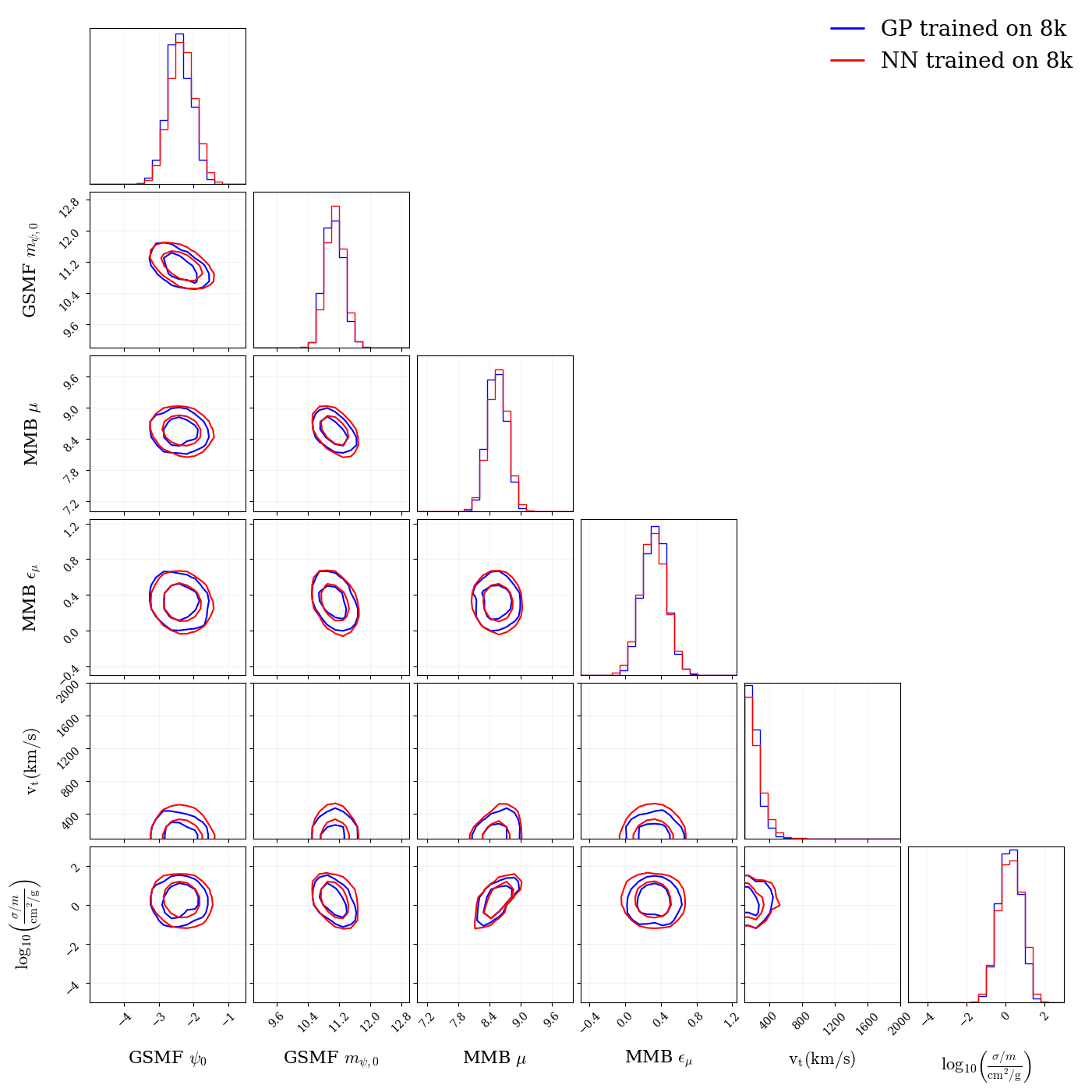}
\caption{Corner plot comparing the posterior distributions of the SIDM model parameters obtained from MCMC analyses using GP and NN interpolators trained on the same 8000-point library. The contours denote the 68\% and 95\% credible regions. The close agreement between the two sets of contours indicates that replacing the GP with an NN does not significantly alter the inferred parameter constraints.}
  \label{fig:mcmc_gp_nn_8k_sidm}
\end{figure*}

In Fig.~\ref{fig:mcmc_gp_nn_8k_sidm}, we show posterior distributions from MCMC runs using GPs and NNs trained on 8000 points. This plot demonstrates that the NNs produce a very similar posterior distribution to GPs. 

\section{Phenomenological model}
In addition to the SIDM model, we apply our accelerated
statistical-analysis pipeline to a second astrophysical model:
the phenomenological model used by
\hyperlink{cite.Agazie_2023}{Agazie2023}. As discussed there,
a direct treatment of black hole binary evolution, including
environmental effects, can introduce many additional free
parameters. For this reason, they adopt a phenomenological
model in which the overall environmental contribution to the
binary evolution is represented by a double power law. This
formulation captures the essential dynamics without introducing
an excessive number of model parameters.

Evolving a binary system involves calculating the orbital decay rate, \(da/dt\), where \(a\) is the binary separation. This decay rate determines the dynamics of the system, and in turn, the GW emission. In the phenomenological model, this quantity is given by
\begin{equation}
    \left| \frac{da}{dt} \right|_{\mathrm{phenom}}
    = H_a \left( \frac{a}{a_c} \right)^{1 - \nu_{\mathrm{inner}}}
    \left( 1 + \frac{a}{a_c} \right)^{\nu_{\mathrm{inner}} - \nu_{\mathrm{outer}}}.
\end{equation}
Here, \(a_c\) is the critical separation, which they set to 100 parsecs. One of the indices of the power law, \(\nu_\mathrm{outer}\), is fixed to +2.5. \(\nu_\mathrm{inner}\) is a free parameter. \(H_a\) is the normalisation factor, which is computed by imposing the condition:
\begin{equation}
    \tau_f = \int_{a_{\mathrm{init}}}^{a_{\mathrm{isco}}} \left( \frac{da}{dt} \right)^{-1} da.
\end{equation}
Here, \(a_\mathrm{init}\) is the initial separation, which they set to \(10^3\) parsecs, \(a_\mathrm{isco}\) is the innermost stable circular orbit, which is three times the Schwarzschild radius, and \(\tau_f\) is the hardening time, which is the other free parameter.

In addition to these two parameters, the phenomenological model also
includes the four population parameters that were varied in the SIDM
analysis: the GSMF parameters \(\psi_0\) and \(m_{\psi,0}\), and the
MMB-relation parameters \(\mu\) and \(\epsilon_\mu\).
In this analysis, we adopt
uniform priors for all six phenomenological-model parameters.
These parameters and their priors are summarised in
Table~\ref{table:phenom_params}.

\begin{table}[htbp]
    \centering
    {
    \renewcommand{\arraystretch}{1.5}
    \begin{tabular}{|c|c|}
        \hline
        \textbf{Model parameter} & \textbf{Priors} \\
        \hline
        phenom $\tau_f$        & $\mathcal{U}(0.1, 11.0)$ Gyr \\
       phenom $\nu_{\mathrm{inner}}$   & $\mathcal{U}(-1.5, 0.0)$ \\
        \hline
        GSMF $\psi_0$        & $\mathcal{U}(-3.5, -1.5)$ \\
       GSMF $m_{\psi,0}$    & $\mathcal{U}(10.5, 12.5)$ \\
        \hline
       MMB $\mu$           & $\mathcal{U}(7.6, 9.0)$ \\
        MMB $\epsilon_\mu$  & $\mathcal{U}(0.0, 0.9)$ dex \\
        \hline
    \end{tabular}
    }
\caption{Priors for the phenomenological model parameters.}
\label{table:phenom_params}
\end{table}

\begin{figure}[h!]
  \centering
  \includegraphics[width=0.45\textwidth]{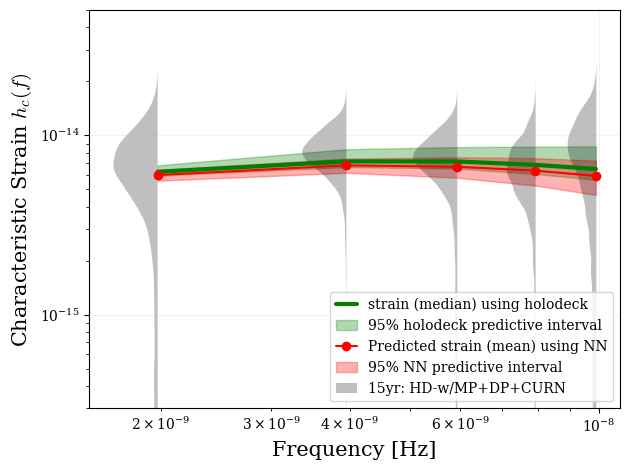}
  \includegraphics[width=0.45\textwidth]{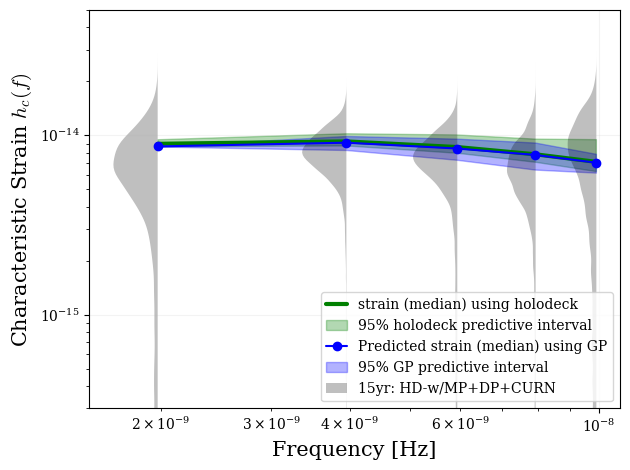}
  \caption{Comparison of interpolator predictions with direct
\texttt{holodeck} simulations for the phenomenological model, using the
same plotting convention as in
Figs.~\ref{fig:gp_holodeck_strain_sidm} and
\ref{fig:nn_holodeck_strain_sidm}. The top panel shows the NN prediction
at the maximum-posterior parameter values obtained from the NN-based
MCMC analysis, while the bottom panel shows the GP prediction at the
maximum-posterior parameter values obtained from the GP-based MCMC
analysis. In each panel, the solid red or blue curve shows the
interpolator-predicted median characteristic-strain spectrum, and the
corresponding shaded region shows the nominal 95\% predictive interval.
The solid green curve shows the median spectrum from direct
\texttt{holodeck} realisations at the same parameter values, and the
green shaded region shows the corresponding central 95\% range. Both the
NN and GP predictions agree well with the direct \texttt{holodeck}
simulations under this maximum-posterior validation check.}
  \label{fig:gp_nn_holodeck_strain_phenom}
\end{figure}

\hyperlink{cite.Agazie_2023}{Agazie2023} generated a 2000-point library to train the GPs, which they used in the MCMC generation. We replaced the GPs in this process with NNs. As in the SIDM case, where we checked whether the GPs are sufficiently trained, we perform a similar test for the phenomenological model. We first check by training the GPs on a 2000-point training dataset, and then, we also train NNs on the same dataset to check if the predictions of the GPs and NNs agree with the simulated spectra at the respective maximum posterior parameter points. We show in Fig.~\ref{fig:gp_nn_holodeck_strain_phenom} that they do agree for both the GPs and the NNs.

\begin{figure*}[htb]
  \centering
  \includegraphics[width=0.9\textwidth]{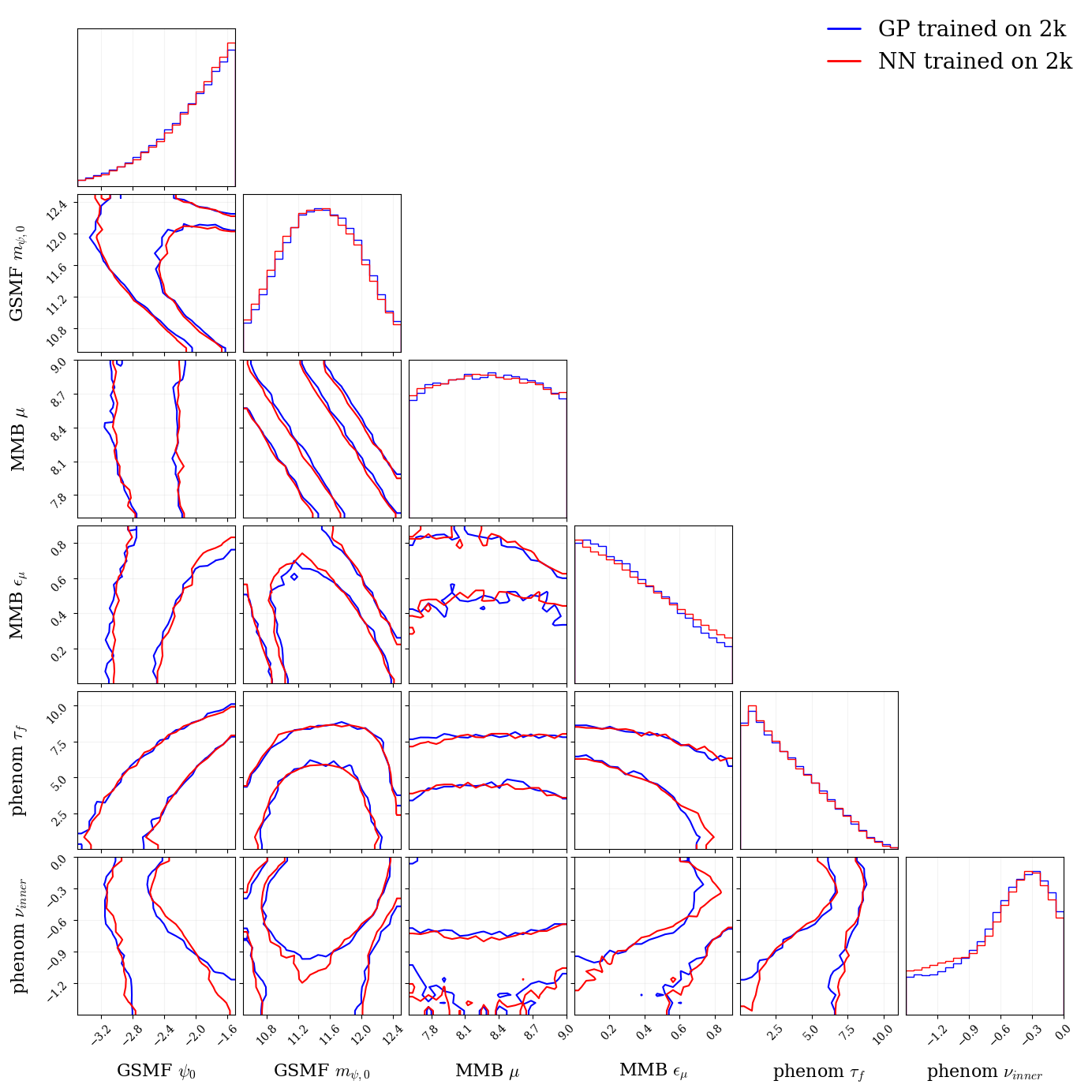}
  \caption{Posterior distributions for the phenomenological model parameters obtained
  using GP and NN interpolators. Contours show the 68\% and 95\% credible regions.}
  \label{fig:mcmc_gp_nn_2k_phenom}
\end{figure*}

For the phenomenological model, training the GPs on a 2000-point dataset took 140.4 minutes. We then used these GPs within the MCMC analysis to generate posterior samples. The posterior distribution of the model parameters is presented in Fig.~\ref{fig:mcmc_gp_nn_2k_phenom}. These results agree well with the results shown by \hyperlink{cite.Agazie_2023}{Agazie2023} in their Fig.~9 (blue lines for Phenom+Uniform).

To train the NN on the median values, we used
three hidden layers with 8, 16, and 8 neurons, respectively.
A NN hyperparameter selection process used here was similar to the one described in Sec.~\ref{subsub:hyperparam_nn}.
To mitigate overfitting, we applied L2 regularisation with
a regularisation parameter of 0.001 to all layers.
This hyperparameter was tuned by monitoring overfitting in the validation loss.
We trained the
NN for up to 1000 epochs, using early stopping based on
the validation loss with \texttt{patience}=100. The minimum
validation loss was reached at epoch 853.

For the NN trained on the standard
deviations, we used the same architecture and training setup.
The minimum validation loss in that case was reached at epoch
424.

We then proceeded to produce the MCMC chains and obtained the posterior distribution. We present this as a corner plot in Fig.~\ref{fig:mcmc_gp_nn_2k_phenom}. 
The NN posteriors are nearly indistinguishable from the GP posteriors. The median values with the  16th and 84th percentiles of the posterior distributions are reported in Table \ref{tab:phenom_posterior_median}.

\begin{table}
\centering
\renewcommand{\arraystretch}{1.6}
\begin{tabular}{|l|r|r|}
\hline
Parameter & GP & NN \\
\hline
GSMF $\psi_0$ & $-1.98^{+0.34}_{-0.59}$ & $-1.96^{+0.33}_{-0.59}$ \\
GSMF $m_{\psi, 0}$ & $11.50^{+0.50}_{-0.49}$ & $11.47^{+0.50}_{-0.49}$ \\
MMB $\mu$ & $8.30^{+0.46}_{-0.46}$ & $8.30^{+0.46}_{-0.46}$ \\
MMB $\epsilon_{\mu}$ & $0.32^{+0.31}_{-0.22}$ & $0.33^{+0.33}_{-0.23}$ \\
phenom $\tau_f$ & $2.89^{+3.05}_{-2.04}$ & $2.84^{+3.07}_{-1.99}$ \\
phenom $\nu_{\mathrm{inner}}$ & $-0.49^{+0.31}_{-0.52}$ & $-0.50^{+0.32}_{-0.57}$ \\
\hline
\end{tabular}
\caption{Posterior constraints on the phenomenological-model parameters obtained from MCMC analyses using GP and NN interpolators. For each parameter, we report the median posterior value and the corresponding 16th and 84th percentiles.}
\label{tab:phenom_posterior_median}
\end{table}

In our own implementation, the time for MCMC generation using GPs was 
129.2 minutes.
 Similarly, for MCMC using NNs for the same number of samples, the total wall-clock time was 
37.5 minutes.

These training times for the GPs and NNs for the phenomenological model are also summarised in Table \ref{table:time}. The process of training the NNs was \(45.3\) times faster than training the GPs. 
We also obtained a speed-up factor of 3.5
 in MCMC runs with the NNs.

\section{\label{sec:results}Results}

In this section, we compare GP and NN interpolators for
both the SIDM and phenomenological models in terms of
training time, predictive performance, and posterior
recovery.

\subsection{Training time}

The training-time comparison is summarised in
Table~\ref{table:time}. For the SIDM model, GP training took 147.4 times as long as NN training, while for the
phenomenological model, the GP training took 45.3 times as long as NN training. The larger gain in the SIDM case reflects the
greater cost of GP training for the larger 8000-point
library.

\subsection{Predictive performance}

\begin{figure}[ht]
  \centering
  \includegraphics[width=0.45\textwidth]{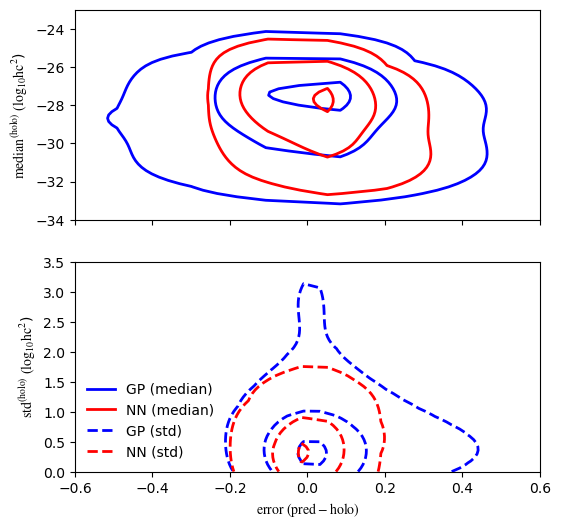}
  \caption{GP and NN predictive errors for the SIDM model.
  Both interpolators are trained on 8000 training points.
  The error is defined as the predicted value minus the
  \texttt{holodeck}-simulated value for the median
  (top panel) and standard deviation (bottom panel) of
  the strain spectrum in the test dataset. We show the
  20th, 50th, and 90th percentiles of these errors. GP
  errors are shown in blue and NN errors in red.}
  \label{fig:gp_nn_error_sidm}
\end{figure}

\begin{figure}[ht]
  \centering
  \includegraphics[width=0.45\textwidth]{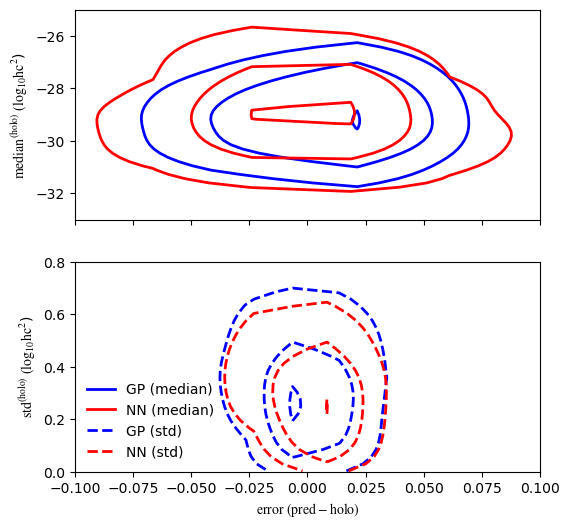}
  \caption{Same as Fig.~\ref{fig:gp_nn_error_sidm}, but for
  the phenomenological model with 2000 training points.}
  \label{fig:gp_nn_error_phenom}
\end{figure}

To assess predictive performance, we compare NNs and GPs
trained on the same training sets and evaluated on the
same test sets. Following Fig.~6 of
\hyperlink{cite.Agazie_2023}{Agazie2023}, we compare the
predictions of each interpolator with the corresponding
\texttt{holodeck} values in the test set and plot the
resulting predictive errors in
Figs.~\ref{fig:gp_nn_error_sidm} and
\ref{fig:gp_nn_error_phenom}.

Figure~\ref{fig:gp_nn_error_sidm} shows that, for the
SIDM model, the NNs trained on 8000 points outperform
the GPs. For the phenomenological model, shown in
Fig.~\ref{fig:gp_nn_error_phenom}, the two methods
perform similarly, with the GP performing marginally
better for the median prediction. A likely reason is that
the phenomenological-model training set contains only
2000 points, which is less favourable for NN training.

The predictive errors are also generally smaller for the phenomenological
model than for the SIDM model. This may reflect the larger scatter in
the SIDM strain realisations, which makes the interpolation problem more
demanding. Consistent with this interpretation,
Figs.~\ref{fig:gp_holodeck_strain_sidm} and
\ref{fig:nn_holodeck_strain_sidm} show that, at the
maximum-posterior parameter values, increasing the SIDM training-library
size improves the GP agreement with the direct \texttt{holodeck}
simulations, while the NN interpolator satisfies the median-containment
validation criterion for each training-set size considered.

\subsection{Posterior recovery and runtime}

We perform Bayesian inference using MCMC, with the
likelihood evaluated as described in Sec.~3.5 of
\hyperlink{cite.Agazie_2023}{Agazie2023}. For each
sampled parameter combination, the interpolator must
predict the distributions of the median and standard
deviation that enter the likelihood calculation.

For the SIDM model, the total MCMC wall-clock time was
2609.7 minutes when using GPs. For the same number of
samples and the same computational setup, the NN-based
analysis required only 39.6 minutes, corresponding to a
speed-up factor of \(65.9\). As shown in
Fig.~\ref{fig:mcmc_gp_nn_8k_sidm}, the NN-based
posteriors are almost identical to those obtained with
the GP-based analysis.

For the phenomenological model, the speed-up is smaller
because the GPs are trained on a smaller 2000-point
library and are therefore less expensive to evaluate. The
GP-based analysis required 129.2 minutes, whereas the
NN-based analysis required 37.5 minutes, corresponding
to a speed-up factor of \(3.5\). As shown in
Fig.~\ref{fig:mcmc_gp_nn_2k_phenom}, the resulting
posterior distributions are again very similar,
indicating that the NN interpolator successfully
reproduces the GP-based constraints.

Overall, these results show that replacing GPs with NNs substantially
reduces the computational cost of the inference pipeline, including both
interpolator training and MCMC sampling, while leaving the inferred
posterior constraints essentially unchanged. The timing comparison is
summarised in Table~\ref{table:time}.

\begin{table}[htbp]
    \centering
    \renewcommand{\arraystretch}{1.5}
    \begin{tabular}{|c|c|c|c|c|}
        \hline
        \multirow{2}{*}{\textbf{Model}} &
        \multirow{2}{*}{\textbf{Process}} &
        \(\mathbf{t_{\mathrm{GP}}}\) &
        \(\mathbf{t_{\mathrm{NN}}}\) &
        \textbf{Ratio} \\
        & & \textbf{(minutes)} & \textbf{(minutes)} & \(\mathbf{\left(t_{\mathrm{GP}}/t_{\mathrm{NN}}\right)}\) \\
        \hline
        \multirow{4}{*}{\rotatebox[origin=c]{90}{SIDM}}
            & Library generation     & 562.4  & 843.6 & 0.7   \\
            & Interpolator training  & 1976.5 & 13.4  & 147.4 \\
            & MCMC generation        & 2609.7 & 39.6  & 65.9  \\
        \cline{2-5}
            & Total                  & 5148.6 & 896.6 & 5.7   \\
        \hline
        \hline
        \multirow{4}{*}{\rotatebox[origin=c]{90}{Phenom}}
            & Library generation     & 37.4   & 56.1  & 0.7   \\
            & Interpolator training  & 140.4  & 3.1   & 45.3  \\
            & MCMC generation        & 129.2  & 37.5  & 3.5   \\
        \cline{2-5}
            & Total                  & 307.0  & 96.7  & 3.2   \\
        \hline
    \end{tabular}
    \caption{Wall-clock times for library generation,
    interpolator training, and MCMC sampling with GP and
    NN interpolators for the SIDM and phenomenological
    models. The training libraries were generated in
    parallel on 128 CPU cores and contain 8000 parameter
    points for SIDM and 2000 for the phenomenological
    model. For the NN runs, the reported
    library-generation times also include the generation
    of validation sets containing 4000 points for SIDM
    and 1000 points for the phenomenological model. MCMC
    timings are reported for approximately \(10^5\)
    samples.}
    \label{table:time}
\end{table}

The total wall-clock time for the full \texttt{holodeck}
analysis pipeline in the SIDM case is reduced from
5148.6 minutes (3.6 days) with GPs to 896.6 minutes
(0.6 days) with NNs, corresponding to an overall
speed-up factor of \(5.7\). For the phenomenological
model, the total pipeline time is reduced from
307.0 minutes (5.1 hours) to 96.7 minutes (1.6 hours),
corresponding to an overall speed-up factor of \(3.2\).

\section{\label{sec:conclusion}Conclusion}

We have investigated probabilistic NNs
as replacements for GP interpolators
in a Bayesian pulsar-timing-array inference pipeline for
the nanohertz GWB. We
considered two source models implemented in the
\texttt{holodeck} framework: a six-parameter
SIDM model and a
six-parameter phenomenological environmental model.

Our main result is that probabilistic NNs can replace GP
interpolators in this setting without degrading the
inferred astrophysical constraints, while substantially
reducing the total computational cost. For the SIDM
model, the total pipeline wall-clock time was reduced
from 3.6 days to 0.6 days, corresponding to an overall speed-up factor
of \(5.7\). For the phenomenological model, the total
pipeline wall-clock time was reduced from 5.1 hours to 1.6 hours, corresponding
to an overall speed-up factor of \(3.2\). The largest
gain comes from eliminating the GP training bottleneck,
with further reductions in the MCMC wall-clock time.

These computational gains do not come at the expense of
inference quality. 
For the SIDM model, the NN interpolators satisfy our
maximum-posterior validation check. The directly simulated
 \texttt{holodeck} median lies within the NN 95\% predictive interval
in each of the five PTA frequency bins.
By contrast, the corresponding GP predictions in Fig.~\ref{fig:gp_holodeck_strain_sidm} do not satisfy this median-containment criterion, even though the GP- and NN-based posteriors remain in close agreement.
For the phenomenological model, the NN and
GP interpolators perform similarly. In both cases, the
posterior distributions obtained with NNs are in very
good agreement with those obtained with GPs.

More elaborate emulators that model the full strain-realization distribution are useful when the distribution itself is the target. For the likelihood-level comparison performed here, however, the relevant requirement is more limited: accurate recovery of the median, predictive uncertainty, and resulting parameter posteriors within the existing \hyperlink{cite.Agazie_2023}{Agazie2023} inference framework.

The main practical advantage of the NN approach is that it removes the GP training bottleneck for large training libraries, thereby making the pipeline more scalable for higher-dimensional and more computationally expensive astrophysical models. Although we demonstrate the emulator using the NANOGrav 15-year data set, the replacement of GP interpolation by probabilistic neural-network emulation is, in principle, independent of the particular PTA data set and can be applied to future NANOGrav, Parkes Pulsar Timing Array, European Pulsar Timing Array, Chinese Pulsar Timing Array, and International Pulsar Timing Array analyses. Probabilistic NNs, therefore, offer a robust and efficient replacement for GP interpolation in PTA GWB inference.

\FloatBarrier
\section*{Acknowledgments}
We gratefully acknowledge support from the Marsden Fund Council grant MFP-UOA2131 from New Zealand Government funding, managed by the Royal Society Te Ap\={a}rangi. We thank Guilhem Lavau and Florent Leclercq for helpful discussions. We also acknowledge the University of Canterbury Research Cluster facilities for providing computational resources that significantly improved the efficiency of our computations (\href{https://doi.org/10.18124/CANTERBURYNZ-UCRCH}{DOI:10.18124/CANTERBURYNZ-UCRCH}, RRID:SCR\_027870).

\bibliographystyle{mnras}
\bibliography{mainbib}

\end{document}